\documentclass[reqno]{amsart}

\newcommand{\Rm}{\mathbb{R}}

\newcommand{\be}{\begin{equation}}
\newcommand{\ee}{\end{equation}}
\newcommand{\va}{\varphi}

\newcommand{\pp}{\partial}

\usepackage{amsmath}
\usepackage[foot]{amsaddr}
\usepackage{amsthm,amssymb,mathrsfs}
\usepackage{graphicx}
\usepackage{bm}
\usepackage{color}

\newtheorem{thm}{Theorem}[section]
\newtheorem{lem}[thm]{Lemma}

\newtheorem{defn}[thm]{Definition}
\theoremstyle{remark}\newtheorem{rmk}[thm]{Remark}

\title[]{Born series for the photon diffusion equation perturbing the Robin boundary condition}

\author[]{Manabu Machida}
\address{Institute for Medical Photonics Research, 
Hamamatsu University School of Medicine, 
Hamamatsu 431-3192, Japan}
\email{machida@hama-med.ac.jp}

\author[]{Gen Nakamura}
\address{Department of Mathematics, Hokkaido University, 
Sapporo 060-0810, Japan}
\email{nakamuragenn@gmail.com}

\date{\today}

\begin{document}

\begin{abstract}
The photon diffusion equation is solved making use of the Born series for 
the Robin boundary condition. We develop a general theory for arbitrary 
domains with smooth enough boundaries and explore the convergence. The 
proposed Born series is validated by numerical calculation in the 
three-dimensional half space. It is shown that in this case the Born series 
converges regardless the value of the impedance term in the Robin boundary 
condition.
\end{abstract}

\maketitle

\section{Introduction}

Diffusion is often seen in different subfields of science and engineering. In particular, light propagation in turbid media such as biological tissue is governed by the diffusion equation except near sources and boundaries \cite{Ishimaru78}. There are scattering and absorption in the medium and they are characterized by the diffusion and absorption coefficients in the diffusion equation, or the photon diffusion equation emphasizing the existence of the absorption term. In addition to its importance in natural science, diffusion in random media has been utilized in medicine \cite{Yodh-Chance95}. Diffuse optical tomography (DOT) is a near-infrared version of X-ray computed tomography \cite{Boas-etal2001}, for which inverse problems are to determine the diffusion coefficient, the absorption coefficient, or both from boundary measurements \cite{Arridge99,Arridge-Schotland09}. Brain activity has been investigated by near-infrared spectroscopy from boundary measurements of diffuse light \cite{Hoshi16}.

At the depth of about ten times the transport mean free path, the energy density of light, which is governed by the Maxwell equations, starts to obey the diffusion equation via the mesoscopic regime of the radiative transport equation \cite{Ryzhik-Papanicolaou-Keller96,Yoo-Liu-Alfano90}. Therefore for highly scattering media such as biological tissue, the diffusion regime becomes dominant. Hence it is common to assume that the diffusion regime spans the whole domain including the boundary. Then the energy density of light in the medium is obtained as the solution to the photon diffusion equation with the Robin boundary condition.

In this paper, we consider the Born sequence for the Robin boundary condition and derive the solution to the diffusion equation as a series. The convergence of the Born series is tested when the spatial domain is a three dimensional half space. More precisely, for a diffusion equation with homogeneous diffusion coefficient and absorption coefficient given in the half space over some finite time interval with the Robin boundary condition, we tested the convergence of the Born series for the Poisson kernel when we treat the Robin boundary condition as perturbation of the Neumann boundary condition. A striking result given later in Sec.~\ref{half} (see Remark \ref{half:doublefactorial}) is that this Born series converges even when the homogeneous impedance term of the Robin boundary condition is not small.

The rest of this paper is organized as follows. In Sec.~\ref{ana} we will discuss the efficiency of the so-called extrapolated boundary condition, which has been used in the study of optical tomography. This boundary condition was introduced to give an approximate solution in a concise way for the initial boundary value problem for the aforementioned diffusion equation with the Robin boundary condition. We show that the efficiency of this boundary condition is limited, which led us to our study given in this paper. Section \ref{general} is devoted to a general study of the Born approximation.  Then based on this general study, we define in Sec.~\ref{scheme} the Born approximation for the Poisson kernel in the half space and a slab domain over some finite time interval. Further, we discuss its convergence of the Born approximation for the Poisson kernel in the half space over some finite time interval in Sec.~\ref{half}. In Sec.~\ref{numerical}, we tested the numerical performance of the Born approximation for the Poisson kernel in the half space over some finite time interval. The last section is for concluding remarks. Appendices \ref{alt} through \ref{specialfunctions} give some supplementary arguments and facts which are better to be separated from the main part of this paper to clarify the points of arguments.

\section{Analytical solution and extrapolated boundary}
\label{ana}

Let us consider the domain $\Omega=\Rm^3_+$, where $\Rm^3_+=\left\{x\in\Rm^3;\,x_3>0\right\}$. The boundary, i.e., the $x_1$-$x_2$ plane, is denoted by $\pp\Omega$. We will find an expression for $u$ which satisfies
\be
\left\{\begin{array}{lll}
\left(\pp_t-\gamma\Delta+b\right)u=0,
&\qquad(x,t)\in\Omega_T,
\\
\gamma\pp_{\nu}u+\beta u=\delta(x_1-y_1)\delta(x_2-y_2)\delta(t-s),
&\qquad(x,t)\in\pp\Omega_T,
\\
u=0,
&\qquad x\in\Omega,\quad t=0,
\end{array}\right.
\label{ana:de}
\ee
where in the Robin boundary condition, $\pp_{\nu}=\nu\cdot\nabla$ with $\nu$ the unit normal of $\partial\Omega$ directed into the exterior of $\Omega$. 

Considerable efforts have been paid to derive concise solution formulae for the diffusion equation \cite{Arridge-Cope-Delpy92,Hielscher-Jacques-Wang-Tittel95,Martelli-etal}. Among such efforts, the extrapolated boundary is a {\em fudged-up} boundary (Chapter 5 in the book by Duderstadt and Hamilton \cite{Duderstadt-Hamilton}) placed in an infinite medium obtained by removing the true boundary. Although it is not easy to mathematically justify the validity of the extrapolated boundary condition, this boundary condition has been successfully used for light propagation in biological tissue \cite{Ayyalasomayajula-Yalavarthy13,Kienle-etal98,Patterson-Chance-Wilson89,Schweiger-etal95}. 

The diffusion equation with the extrapolated boundary condition is described as the following initial value problem for $u_{\rm EBC}(x,t)$: 
\begin{equation}\label{eq:::uEBC}
\left\{\begin{array}{lll}
\left(\pp_t-\gamma\Delta+b\right)u_{\rm EBC}=
\delta(x_1-y_1)\delta(x_2-y_2)
\left[\delta(x_3)-\delta(x_3+2\ell)\right]\delta(t-s),
\\
\qquad\qquad\qquad\qquad\qquad\qquad\qquad\qquad\qquad\qquad(x,t)\in\Rm^3\times(0,T),
\\
u_{\rm EBC}=0,
\quad x\in\Rm^3,\quad t=0.
\end{array}\right.
\end{equation}
Here the ratio $\ell=\gamma/\beta$ is called the extrapolation distance. $u_{\rm EBC}$ restricted to $\Omega$ will be considered to approximate the solution $u$ of \eqref{ana:de}. When $\ell$ is close to $0$, the boundary $\partial\Omega$ is almost purely absorbing, and the purely reflecting boundary is achieved in the limit $\ell\to\infty$. We remark that sometimes the source is placed inside the medium with the source term given by $\delta(x_1-y_1)\delta(x_2-y_2)\left[\delta(x_3-d)-\delta(x_3+2\ell-d)\right]\delta(t-s)$, where $d$ is about the transport mean free path \cite{Patterson-Chance-Wilson89}.

We briefly examine the performance of approximating  $u$ by $u_{\rm EBC}(x,t)$ restricting to $\Omega$. In Appendix \ref{alt} we explicitly calculate the solution to (\ref{ana:de}) in the half space. We put
\be
y_1=y_2=s=0.
\label{ana:set0}
\ee
Then the exact solution to (\ref{ana:de}) at $x_1=x_2=0$ for $x_3>0, t\ge0$ is given by
\be
u(x,t)=u(x_3,t)=\frac{2e^{-bt}}{4\pi\gamma t}\left[
\frac{e^{-\frac{x_3^2}{4\gamma t}}}{\sqrt{4\pi\gamma t}}-
\frac{\beta}{2\gamma}e^{\frac{\beta}{\gamma}(x_3+\beta t)}
\mathop{\mathrm{erfc}}\left(\frac{x_3+2\beta t}{\sqrt{4\gamma t}}\right)
\right],
\label{ana:solexact}
\ee
where the complementary error function $\mathop{\mathrm{erfc}}(\xi)$, 
$\xi\in\Rm$ is defined as
\[
\mathop{\mathrm{erfc}}(\xi)=
\frac{2}{\sqrt{\pi}}\int_{\xi}^{\infty}e^{-s^2}\,ds.
\]
Furthermore we obtain
\be
u_{\rm EBC}(x_3,t)=
\frac{e^{-bt}}{(4\pi\gamma t)^{3/2}}\left(
e^{-\frac{x_3^2}{4\gamma t}}-e^{-\frac{(x_3+2\gamma/\beta)^2}{4\gamma t}}
\right).
\label{ana:solEBC}
\ee

Let us numerically compare $u$ and the restriction of $u_{\rm EBC}$ to $\Omega$. First of all noticing $\mathop{\mathrm{erfc}}(\xi)=\frac{1}{\sqrt{\pi}}e^{-\xi^2}\left(\xi^{-1}+O(\xi^{-3})\right)$ for large $\xi$, we have
\[
\begin{array}{ll}
\left|\frac{u_{\rm EBC}(x_3,t)-u(x_3,t)}{u(x_3,t)}\right|
&=
\frac{1+e^{-\frac{x_3+\gamma/\beta}{\beta t}}-
\frac{\beta}{\gamma}\sqrt{4\gamma t}\left(\xi^{-1}+O(\xi^{-3})\right)}
{2-\frac{\beta}{\gamma}\sqrt{\frac{\gamma t}{\pi}}
\left(\xi^{-1}+O(\xi^{-3})\right)},
\end{array}
\]
where $\xi=(x_3+2\beta t)/\sqrt{4\gamma t}$. Therefore we obtain $\lim_{x_3\to\infty}\left|(u_{\rm EBC}-u)/u\right|=1/2\neq0$ although $\lim_{t\to\infty}\left|(u_{\rm EBC}-u)/u\right|=\lim_{\beta\to\infty}\left|(u_{\rm EBC}-u)/u\right|=0$.

Next we set
\be
\gamma=0.06\,{\rm mm}^2/{\rm ps},\qquad b=0.001\,{\rm ps}^{-1},\qquad
T=4\,{\rm ns},\qquad x_3=20\,{\rm mm}.
\label{ana:para}
\ee
In Fig.~\ref{fig:ebc} below, we compare $u(x_3,t)$ in (\ref{ana:solexact}) and 
$u_{\rm EBC}(x_3,t)$ in (\ref{ana:solEBC}). When $\beta$ is small, the agreement is not good. As $\beta$ becomes larger, $u_{\rm EBC}$ approaches the exact solution $u(x_3,t)$.

\begin{figure}[ht]
\includegraphics[width=0.3\textwidth]{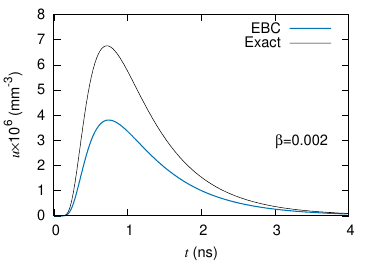}
\includegraphics[width=0.3\textwidth]{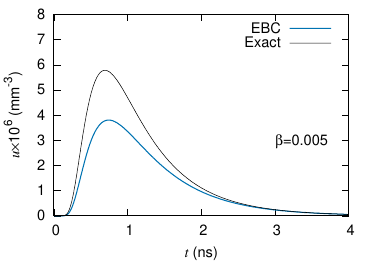}
\includegraphics[width=0.3\textwidth]{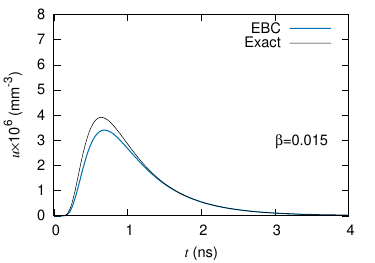}
\caption{
The energy density $u$ is plotted at $x_3=20\,{\rm mm}$ as a function of $t$ for, from the left to right, $\beta=0.002\,{\rm mm}/{\rm ps}$, $=0.005\,{\rm mm}/{\rm ps}$, and $=0.015\,{\rm mm}/{\rm ps}$, respectively. In each panel, $u(x_3,t)$ and $u_{\rm EBC}(x_3,t)$ are compared.
}
\label{fig:ebc}
\end{figure}

\section{General theory for Born series}
\label{general}

In this section a general scheme is given to define the Born series for the initial boundary value problem for the diffusion equation with the Robin boundary condition. The impedance term (i.e., $\beta u$ in (\ref{ana:de})) in the Robin 
boundary condition is considered as a perturbation for the Born series. 

Throughout this section let $\Omega$ be a domain in $\Rm^n$ $(n=2, 3)$ and $\pp\Omega$ be the boundary of $\Omega$ which is of $C^2$ class. For simplicity of description we only describe our scheme for $n=3$. We define
\[
\Omega_T=\Omega\times(0,T),\qquad\pp\Omega_T=\pp\Omega\times(0,T),
\qquad T>0.
\]
Let $\gamma=(\gamma_{ij})$ and $b$ be the diffusion coefficient and the absorption coefficient which are bounded measurable in $\Omega$, i.e., $\gamma,\,b\in L^\infty(\Omega)$. We assume that there exists a positive constant $\delta$ such that 
\be\label{assumption}
\left\{
\begin{array}{l}
b\ge\delta,\\
\displaystyle\sum_{i,j=1}^3\gamma_{ij}(x)\xi_i\xi_j\ge\delta\displaystyle\sum_{i=1}^3\xi_i^2\,\,\text{for any $\xi=(\xi_1,\xi_2,\xi_3)\in{\mathbb R}^3$}
\end{array}
\right.
\ee 
almost everywhere in $\Omega$. Now we consider the following initial boundary value problem for the diffusion equation for the energy density $u(x,t)$:
\be
\left\{\begin{array}{lll}
\left(\pp_t-\nabla\cdot\gamma\nabla+b\right)u=f,
&\qquad(x,t)\in\Omega_T,
\\
\gamma\partial_\nu u+\beta u=g,
&\qquad(x,t)\in\pp\Omega_T,
\\
u=0,
&\qquad x\in\Omega,\quad t=0,
\end{array}\right.
\label{de1}
\ee
where $f=f(x,t)$ is the internal source, $g=g(x,t)$ is the boundary source and $\beta$ is a positive bounded measurable function on $\partial\Omega$, i.e., $\beta\in L^\infty(\partial\Omega)$. For the simplicity of description we assume $\gamma=\gamma(x) I$ with scalar function $\gamma(x)\in L^\infty(\Omega)$ abusing the notation $\gamma$ and the $3\times3$ identity matrix $I$.

\begin{rmk}
We can include the incident beam $h(x)$ in the initial condition of (\ref{de1}). By Duhamel's principle, however, it can reduce to the case $h=0$.
\end{rmk}

In order to give the definition of the weak solution to \eqref{de1}, we first introduce $L^2$-Sobolev spaces and related function spaces. Let $H^1(\Omega)$ be the real $L^2$ Sobolev space of order $1$ in $\Omega$ and we denote its dual space by $H^1(\Omega)^*$. Similarly we let $H^{1/2}(\partial\Omega)$ be the real $L^2$ Sobolev space of order $1/2$ on $\partial\Omega$ and denote its dual space by $H^{-1/2}(\partial\Omega)$. Define the trace operator $\Lambda: H^1(\Omega)\ni\psi\mapsto \psi\big|_{\partial\Omega}\in H^{1/2}(\partial\Omega)$. Here we can consider $\Lambda\psi\in H^{-1/2}(\partial\Omega)$ because $H^{1/2}(\partial\Omega)\subset H^{-1/2}(\partial\Omega)$. We will use the pairings $\langle\cdot,\cdot\rangle_\Omega$ and $\langle\cdot,\cdot\rangle_{\partial\Omega}$ for the pairs $(H^1(\Omega), H^1(\Omega)^*)$ and $(H^{1/2}(\partial\Omega), H^{-1/2}(\partial\Omega))$, respectively. Furthermore for any real Hilbert space $E$, $L^2((0,T); E)$ denotes the set of all $E$ valued $L^2$ functions over the time interval $(0,T)$. We will denote its norm by $\Vert\cdot\Vert_{L^2((0,T);E)}$. Throughout the paper, $W((0,T))$ denotes the space defined as
\[
W((0,T)):=\{u=u(t):=u(\cdot, t)\in L^2((0,T);H^1(\Omega)); \partial_t u\in L^2((0,T); H^1(\Omega)^*)\},
\]
equipped with the norm
\[
\|u\|_{W((0,T))}^2:=\int_0^T\left(\|u(t)\|_{H^1(\Omega)}^2+\|\pp_tu(t)\|_{H^1(\Omega)^*}^2\right)\,dt.
\]

We define the weak solution $u$ to \eqref{de1} as follows.

\begin{defn}\label{weak solution}
Let $f=f(t):=f(\cdot,t)\in L^2((0,T); H^1(\Omega))$ and $g=g(t):=g(\cdot,t)\in L^2((0,T); H^{-1/2}(\partial\Omega))$. Then $u\in W((0,T))$ is called the weak solution to \eqref{de1} if it satisfies $u(0)=0$ and
\begin{equation}\label{weak sol1}
\begin{array}{l}
\int_0^T\{\langle\partial_tu,\varphi(t)\rangle_\Omega+
\langle\gamma\nabla u(t),\nabla\varphi(t)\rangle_\Omega+\langle b u(t),\varphi(t)\rangle_\Omega+
\langle\beta \Lambda u(t), \Lambda\varphi(t)\rangle_{\partial\Omega}\}\,dt\\
\qquad\qquad=\int_0^T\langle f(t),\va(t)\rangle_\Omega\,dt+\int_0^T\langle g(t),\Lambda\va(t)\rangle_{\partial\Omega}\,dt
\end{array}
\end{equation}
for any $\varphi=\varphi(t):=\varphi(\cdot,t)\in Z((0,T)):=\{\varphi, \,\partial_t\varphi\in L^2((0,T); H^1(\Omega)),\,\varphi(T)=0\}$.
\end{defn}

\subsection{Operators $A,\,A_0$}${}$
\newline
\indent
Let us consider the following sesquilinear forms:
\begin{equation}\label{eq::sesquilinear forms}
\left\{\begin{array}{ll}
a(v,w):=
\int_{\Omega}(\gamma\nabla v\cdot\nabla w+bvw)+\int_{\pp\Omega}\beta vw,
\\
a_0(v,w):=\int_{\Omega}(\gamma\nabla v\cdot\nabla w+bvw),
\end{array}
\right.
\end{equation}
where $v,w\in H^1(\Omega)$. By using $\gamma,\, b\in L^\infty(\Omega)$, \eqref{assumption} and the positivity of $\beta\in L^\infty(\partial\Omega)$, we can show that $a(v,w),\, a_0(v,w)$ are bounded, symmetric and positive bilinear forms \cite{Wloka}. That is,
\begin{equation}
\label{coercive}
\left\{
\begin{array}{ll}
|a(v,w)|, |a_0(v,w)|\le C_1\Vert v\Vert_{H^1(\Omega)}\Vert w\Vert_{H^1(\Omega)}\,\,\text{\rm (bounded)},
\\
a(v,w)=a(w,v),\,a_0(v,w)=a_0(w,v)\,\,\text{\rm (symmetric)},
\\
a(v,v),\,a_0(v,v)\ge C_2\Vert v\Vert_{H^1(\Omega)}^2\,\,\text{\rm (positive)}
\end{array}
\right.
\end{equation}
for any $v, w\in H^1(\Omega)$ with some positive constants $C_1,\,C_2$ independent of $v, w$. Here we denoted the $H^1(\Omega)$ norm of $v\in H^1(\Omega)$ by $\Vert v\Vert_{H^1(\Omega)}$.
For $v\in H^1(\Omega)$ and $v_0\in H^1(\Omega)$, let $\Psi\in H^1(\Omega)^*$ and $\Psi_0\in H^1(\Omega)^*$ be such that for any $\phi\in H^1(\Omega)$, $\Psi(\phi)=a(v,\phi)$ and $\Psi_0(\phi)=a_0(v_0,\phi)$, respectively. Then define $A$ and $A_0$ by $Av=\Psi$ and $A_0 v_0=\Psi_0$, respectively. From the properties \eqref{coercive}, we see that $A$ and $A_0$ are isomorphisms from $H^1(\Omega)$ to $H^1(\Omega)^*$ \cite{Wloka}.

Now we define $F=F(t)\in L^2((0,T); H^1(\Omega)^*)$ by
\begin{equation}\label{eq::F}
\langle F(t), w\rangle_\Omega=\langle f(t), w\rangle_\Omega+\langle g(t), \Lambda w\rangle_{\partial\Omega},\,\,{\rm a.e.}\,t\in (0,T)
\end{equation}
for any $w\in H^1(\Omega)$. It is easy to show that the norm $\Vert F\Vert_{L^2((0,T); H^1(\Omega)^*)}$ of $F$ has the estimate:
\begin{equation}
\Vert F\Vert_{L^2((0,T); H^1(\Omega)^*)}\le C\big(\Vert f\Vert_{L^2((0,T); H^1(\Omega)^*)}+\Vert g\Vert_{L^2((0,T); H^{-1/2}(\partial\Omega))}\big)
\end{equation}
with some general constant $C>0$.

Since 
$$
\int_0^T\langle\partial_t u(t), \varphi(t)\rangle_\Omega\,dt+\int_0^T\langle u(t),\partial_t\varphi(t)\rangle_\Omega\,dt=
\langle u(T),\varphi(T)\rangle_\Omega-\langle u(0),\varphi(0)\rangle_\Omega
$$
for any $u=u(t):=u(\cdot,t)\in W((0,T))$ and $\varphi=\varphi(t):=\varphi(\cdot,t)\in L^2((0,T); H^1(\Omega))$ such that $\partial_t \varphi\in L^2((0,T); H^1(\Omega))$, the weak solution $u=u(t):=u(\cdot,t)$ to \eqref{de1} in Definition \ref{weak solution} is equivalent to $u\in W((0,T))$ which satisfies
\begin{equation}\label{equivalent}
\partial_t u+Au=F\,\,{\rm in}\,\,L^2((0,T); H^1(\Omega)^*),\,\, u(0)=0\,\,{\rm in}\,\,H^1(\Omega)^*.
\end{equation}
Similarly the definition of the weak solution  $u=u(t):=u(\cdot,t)$ to \eqref{de1} with $\beta=0$ is equivalent to the definition given by
\begin{equation}\label{equivalent0}
\partial_t u+A_0 u=F\,\,{\rm in}\,\,L^2((0,T); H^1(\Omega)^*),\,\, u(0)=0\,\,{\rm in}\,\,H^1(\Omega)^*.
\end{equation}

The fundamental theorem for the well-posedness of the weak solution to either \eqref{de1} or \eqref{de1} with $\beta=0$ is as follows.

\begin{thm}[Theorem 26.1 of \cite{Wloka}]\label{fundamental thm}
Let $f\in L^2((0,T); H^1(\Omega)^*)$ and $g\in L^2((0,T); H^{-1/2}(\partial\Omega))$. Then there exists a unique solution $u=u(t):=u(\cdot,t)\in W((0,T))$ to either \eqref{de1} or \eqref{de1} with $\beta=0$. Further it satisfies $u\in C^0([0,T]; L^2(\Omega))$ and the estimate,
\begin{equation}\label{estimate}
\Vert u\Vert_{W((0,T))}\le C(\Vert f\Vert_{L^2((0,T);H^1(\Omega)^*)}+\Vert g\Vert_{L^2((0,T); H^{-1/2}(\partial\Omega))})
\end{equation}
with some general constant $C>0$.
\end{thm}

Based on this theorem we will define the solution operator $S_0$ as follows.

\begin{defn}\label{S_0}
Define the Green operator $S_0$ of \eqref{equivalent0} by $S_0 f=u$, where $u\in W((0,T))$ is the solution to \eqref{equivalent0} with $g=0$.
\end{defn}

We note by (\ref{estimate}),
\[
\|S_0f\|_{W((0,T))}\le C\|f\|_{L^2((0,T);H^1(\Omega)^*)}
\]
holds for a general constant $C>0$.

\subsection{Born sequence}

Let $B$ be an operator $B: L^2((0,T); H^1(\Omega))\rightarrow L^2((0,T); H^1(\Omega)^*)$ defined by
\begin{equation}\label{operator B}
\begin{array}{ll}
\int_0^T\langle Bv(t), w(t)\rangle_{\Omega}\,dt=\int_0^T\langle\beta \Lambda v(t),\Lambda w(t)\rangle_{\partial\Omega}\,dt
\\
\\
\qquad\qquad{\rm for}\,\,v=v(t):=v(\cdot,t),\,w=w(t):=w(\cdot,t)\in L^2((0,T);H^1(\Omega)).
\end{array}
\end{equation}
Observe that by using the boundedness of the trace operator $\Lambda: H^1(\Omega)\rightarrow H^{1/2}(\partial\Omega)$, we have for any $v,w\in H^1(\Omega)$,
\be
|\langle\beta\Lambda v,\Lambda w\rangle_{\partial\Omega}|\le
C\|\beta\|_{L^{\infty}(\pp\Omega)}\|v\|_{H^1(\Omega)}\|w\|_{H^1(\Omega)}
\ee
with some general constant $C>0$. This immediately implies the estimate for the norm $\Vert B\Vert$ of $B$:
\be\label{estimate B}
\Vert B\Vert\le C\Vert\beta\Vert_{L^\infty(\partial\Omega)}.
\ee

Now we define a Born sequence $u_n,\,\,n\in{\mathbb Z}_+:={\mathbb N}\cup\{0\}$ which satisfy
\be\label{Born sequence}
\begin{array}{ll}
&\left\{
\begin{array}{l}
\frac{d}{dt}u_0+A_0u_0=F,
\\
u_0\big|_{t=0}=0,
\end{array}
\right.
\\
\\
&
\left\{
\begin{array}{l}
\frac{d}{dt}u_n+A_0u_n=-B u_{n-1}+F,\\
u_n\big|_{t=0}=0
\end{array}
\right.
\end{array}
\ee
for $n\in {\mathbb N}$.

\subsection{Convergence}

In this subsection we will prove the convergence of the Born sequence $u_n,\,n\in{\mathbb Z}_+$. To see this define $v_n\,(n=0,1,2,\cdots)$ by 
\be
\left\{\begin{aligned}
&
v_n:=u_n-u_{n-1}=-S_0 B v_{n-1},\quad n=1,2,\cdots
\\
&
v_0=u_0.
\end{aligned}\right.
\label{bornseries}
\ee
Then we have
\[
\begin{array}{ll}
\|v_n\|_{W((0,T))}
&=
\Vert S_0 B v_{n-1}\Vert_{W((0,T))}
\\
&\le
C\Vert B v_{n-1}\Vert_{L^2((0,T); H^1(\Omega)^*)}
\\
&\le
C\Vert B\Vert
\|v_{n-1}\|_{W((0,T))}
\\
&\le
C\|\beta\|_{L^{\infty}(\pp\Omega)}\|v_{n-1}\|_{W((0,T))},
\qquad n=1,2,\cdots
\end{array}
\]
with some general constants $C>0$ which may be different line by line. Therefore the Born series
\[
u_0+(u_1-u_0)+(u_2-u_1)+\cdots+(u_{n+1}-u_n)+\cdots.
\]
and hence the Born sequence $u_n,\,n=0,1,2,\cdots$ converges to a unique $u\in W(0,T)$ if $C\|\beta\|_{L^{\infty}(\pp\Omega)}<1$. From \eqref{Born sequence} and the boundedness of the operators $A_0$, $B$, this implies
\be
\left\{
\begin{array}{l}
\frac{d}{dt}u+A_0 u=-B u+F,\\
u\big|_{t=0}=0.
\end{array}
\right.
\ee
By $Au-A_0 u=Bu$ and the uniqueness of the weak solution to \eqref{de1}, $u$ is the weak solution to \eqref{de1}.

In the rest of this section, by using the above arguments which we have given so far in this section, we will give the existence of Green function for \eqref{de1} with $g=0$ and Poisson kernel for \eqref{de1} with $f=0$, respectively. We also give the convergence of their associated Born sequences to the Schwartz kernels. The Green function and Poisson kernel are the Schwartz kernels \cite{Hormander} of the operators mapping
$S: L^2((0,T); H^1(\Omega)^\ast)\ni f \mapsto u\in L^2((0,T); H^1(\Omega))$ of \eqref{de1} with $g=0$ and $P: L^2((0,T); H^{-1/2}(\partial\Omega))\ni g \mapsto u\in L^2((0,T); H^1(\Omega))$ of \eqref{de1} with $f=0$, respectively. We refer to $S$ and $P$ as the Green operator and Poisson operator for \eqref{de1}, respectively.

The Poisson operator $P$ can be given as $\lim_{\epsilon\rightarrow+0}S^\epsilon$, where $S^\epsilon$ is the Green operator for \eqref{de1} with homogeneous boundary condition and $f=g\otimes\delta_{\partial\Omega_\epsilon}$. Here, $\delta_{\partial\Omega_\epsilon}$ is the Dirac delta function supported on $\partial\Omega_\epsilon$ and $\partial\Omega_\epsilon$ is the boundary of $\Omega_\epsilon=\{x\in\Omega:\,\text{dist}(x,\partial\Omega)>\epsilon\}$ with the distance $\text{dist}(x,\partial\Omega)$ between $x$ and $\partial\Omega$. More precisely $S^\epsilon$ consists of $S^{\epsilon+}$ and $S^{\epsilon-}$ defined over $\Omega_\epsilon$ and $\Omega\setminus\overline{\Omega_\epsilon}$ with the homogeneous boundary condition $(\gamma\partial_\nu+\beta)S^{\epsilon-}=0$ over $\partial\Omega$ and the transimission boundary condition 
$$
S^{\epsilon+}-S^{\epsilon-}=0,\,\,\,\gamma(\partial_\nu S^{\epsilon+}-\partial_\nu S^{\epsilon-})=I
$$
over $\partial\Omega_\epsilon$ with the identity operator $I$ on $L^2((0,T); H^{-1/2}(\partial\Omega_\epsilon))$ (although the situation is slightly different from here but it is essential the same as in Nakamura-Wang \cite{NW} ). Also the $\lim_{\epsilon\rightarrow+0} S^\epsilon$ means that for every $f\in L^2((0,T); H^{-1/2}(\partial\Omega))$, $\lim_{\epsilon\rightarrow+0} S^\epsilon (f\otimes\delta_{\partial\Omega_\epsilon})$ exists in $L^2((0,T);H^1(\Omega))$. Hence it is enough to consider the existence of  the Green function and its Born approximation.

We first show that $S,\,(-S_0B)^j S,\,j=0,1,2,\cdots$ have Schwartz kernels in the space of distribution $\mathcal{D}'(\Omega_T\times\Omega_T)$ defined in $\Omega_T\times\Omega_T$. By Theorem \ref{fundamental thm}, each of these are continuous linear map from
$L^2((0,T); H^1(\Omega)^\ast)$ to $L^2((0,T); H^1(\Omega))$. We refer to this as the $L^2$-type continuity. Based on this we will show that they have Schwartz kernels. Since the further arguments are the same for each of these maps, we only confine our argument to $S$. What we need to show is that $S: C_0^\infty(\Omega_T)\rightarrow\mathcal{D}'(\Omega_T)$ is linear and continuous, where $\mathcal{D}'(\Omega)$ is the space of distribution defined in $\Omega_T$. We refer to this as the distribution-type continuity. Since the linearity of $S$ is clear, we only need to show the continuity of $S$. In order to see this, let $\varphi_\ell\in C_0^\infty(\Omega_T),\,\ell=1,2,\cdots$ be a sequence such that $\text{supp}\,\varphi_\ell\subset K,\,\ell=1,2,\cdots$ for a compact set $K\subset\Omega_T$ and for each $m\in{\mathbb Z}_+$, $\partial_{t,x}^\alpha\varphi_\ell,\,|\alpha|\le m$ go to zero uniformly in $\Omega_T$ as $\ell\rightarrow\infty$, where $\partial_{t,x}^\alpha=\partial_t^{\alpha_0}\partial_{x_1}^{\alpha_1}\cdots\partial_{x_3}^{\alpha_3},\,\alpha=(\alpha_0,\alpha_1,\alpha_2,\alpha_3),\,|\alpha|=\alpha_0+\alpha_1+\alpha_2+\alpha_3$, $\text{supp}\,\varphi_\ell$ denotes the support of $\varphi_\ell$
and $C_0^\infty(\Omega_T)$ is the set of all smooth functions with supports in $\Omega_T$. 
We denote this as $\varphi_\ell\Rightarrow 0\, (\ell\rightarrow\infty)$ and denote by $\mathcal{D}(\Omega_T)$ the topological vector space $C_0^\infty(\Omega_T)$ equipped with the topology induced by the convergence of sequence $\varphi_\ell\Rightarrow 0\,(\ell\rightarrow\infty)$.
Then $\varphi_\ell\Rightarrow 0\,(\ell\rightarrow\infty)$ implies $\varphi_\ell\rightarrow 0,\,\ell\rightarrow\infty$ in $L^2((0,T); H^1(\Omega)^\ast)$ and hence we have that by the $L^2$-type continuity of $S$, $S\varphi_\ell\rightarrow 0,\,\ell\rightarrow\infty$ in $L^2((0,T); H^1(\Omega))$. This immediately gives us the distribution-type continuity of $S$ described as follows: for any fixed $\psi\in C_0^\infty(\Omega_T)$,  
$$
\langle S\varphi_\ell,\psi\rangle=\int_{\Omega_T} (S\varphi_\ell)(x,t)\psi(x,t)\,dx\,dt\rightarrow 0,\,\,\ell\rightarrow\infty,
$$
where $\langle S\varphi_\ell,\psi\rangle$ denotes the pairing between $S\varphi_\ell\in\mathcal{D}'(\Omega_T)$ and $\psi\in C_0^\infty(\Omega_T)$. Then by the Schwartz kernel theorem, $S$ has its unique Schwartz kernel $H(x,t;y,s)\in\mathcal{D}'
(\Omega_T\times\Omega_T)$ such that
\begin{equation}\label{identity for S}
\langle S\varphi, \psi\rangle=\langle H(x,t; y,s),\psi(x,t)\otimes\varphi(y,s)\rangle,\,\,\varphi(y,s),\,\psi(x,t)\in C_0^\infty(\Omega_T).
\end{equation}

Now recall \eqref{bornseries} and the convergence of the Born sequence $u_n=T_n F,\,n=0,1,\cdots$ with $T_n:=\sum_{j=0}^n (-S_0B)^j SF,\,n=0,1,\cdots$, where we take $F$ given by \eqref{eq::F} with $g=0$. Then by \eqref{identity for S} and the denseness of the finite linear combinations of the functions of the form $\psi(x,t)\otimes\varphi(y,s),\,\varphi,\,\psi\in C_0^\infty(\Omega_T)$ in $\mathcal{D}(\Omega_T\times\Omega_T)$ which is defined similarly as $\mathcal{D}(\Omega_T)$, the sequence of Schwartz kernels $H_n^0(x,t;y,s)\in\mathcal{D}'(\Omega_T\times\Omega_T),\,n=0,1,\cdots$ of $T_n,\,n=0,1,\cdots$
converges to $H(x,t;s,y)\in\mathcal{D}'(\Omega_T\times\Omega_T)$ as $n\rightarrow\infty$.

\section{Born approximation for Poisson kernel}
\label{scheme}

Let us consider the following initial boundary value problem for $u$. 
\be
\left\{\begin{array}{lll}
\left(\pp_t-\gamma\Delta+b\right)u=0,
&\qquad(x,t)\in\Omega_T,
\\
\gamma\pp_{\nu}u+\beta u=g
&\qquad(x,t)\in\pp\Omega_T,
\\
u=0,
&\qquad x\in\Omega,\quad t=0,
\end{array}\right.
\label{scheme:DE0}
\ee
where $\Omega$ is either the half space $\Omega={\mathbb R}_+^3:=\{(x_1,x_2, x_3)\in{\mathbb R}^3:\,x_3>0\}$ or slab domain $\Omega=\{(x_1,x_2, x_3)\in{\mathbb R}^3:\,0<x_3<L\}$. We assume that $\gamma$ is a positive constant and $b,\,\beta$ are nonnegative constants. Here, $g=g(x_1,x_2,t)$ is the boundary source. 

We have already shown in Sec.~\ref{general} the existence of Poisson kernel and the convergence of the associated Born series. The aim of this section is to give an explicit form of the Poisson kernel when $\Omega$ is the aforementioned special and simple domains. The half space and slab domain have been used in DOT. For example, the slab geometry was used for fluorescent DOT \cite{Panasyuk-etal08} and DOT for spatially modulated structured light was developed in the half space \cite{Konecky-etal09}. In these studies, the time-independent diffusion equation was used.

\subsection{Poisson kernel}

Let us begin by considering $u_0$ satisfying the following diffusion equation.
\begin{equation}\label{eq:: eq for u0}
\left\{\begin{array}{lll}
\left(\pp_t-\gamma\Delta+b\right)u_0=0,
&\qquad(x,t)\in\Omega_T,
\\
\gamma\pp_{\nu}u_0=g,
&\qquad(x,t)\in\pp\Omega_T,
\\
u_0=0,
&\qquad x\in\Omega,\quad t=0.
\end{array}\right.
\end{equation}
If we have the Poisson kernel $G(x,t;y_1,y_2,s)$ for \eqref{eq:: eq for u0} which is the Schwartz kernel of the operator $S_0$ given in Definition \ref{S_0}, then $u_0(x,t)$ can be given as
\[
u_0(x,t)=\int_{\pp\Omega_T}G(x,t;y_1,y_2,s)g(y_1,y_2,s)\,dy_1dy_2ds.
\]
Below we calculate the Poisson kernel $G(x,t;y_1,y_2,s)$ in the half space and slab domain.

\subsubsection{Half space}

Let us consider the case of the half space, i.e., $\Omega=\Rm^3_+$ and $\pp\Omega=\Rm^2$. The Poisson kernel $G$ satisfies
\be
\left\{\begin{array}{lll}
\left(\pp_t-\gamma\Delta+b\right)G=0,
&\qquad(x,t)\in\Omega_T,
\\
\gamma\pp_{\nu}G=\delta(x_1-y_1)\delta(x_2-y_2)\delta(t-s),
&\qquad(x,t)\in\pp\Omega_T,
\\
G=0,
&\qquad x\in\Omega,\quad t=0.
\end{array}\right.
\label{scheme:Geq}
\ee
Let us introduce $K(x,t;y,s)$ which satisfies
\[
\left\{\begin{array}{lll}
\left(\pp_t-\gamma\Delta+b\right)K=\delta(x-y)\delta(t-s),
&\qquad(x,t)\in\Rm^3\times(0,T),
\\
K=0,
&\qquad x\in\Rm^3,\quad t=0.
\end{array}\right.
\]
We will obtain $K(s,t;y,s)$ by using its Laplace-Fourier transform:
\[
\hat{K}(x_3,y_3)=\hat{K}(x_3;p,q;y,s)=\int_0^{\infty}\int_{\Rm^2}e^{-pt}
e^{-i(q_1x_1+q_2x_2)}K(x,t;y,s)\,dx_1dx_2dt.
\]
Then $\hat{K}$ has to satisfy
\[
-\frac{d^2}{dx_3^2}\hat{K}+\lambda^2\hat{K}=
e^{-ps}e^{-i(q_1y_1+q_2y_2)}\delta(x_3-y_3),\qquad x_3\in\Rm^3
\]
with
\be
\lambda=\sqrt{\frac{b+p}{\gamma}+q\cdot q\,}.
\label{scheme:deflambda}
\ee
The above equation can be solved by the Fourier transform with respect to $x_3$ and  we obtain
\be
\hat{K}(x_3,y_3)=\frac{1}{2\lambda\gamma}
e^{-ps}e^{-i(q_1y_1+q_2y_2)}e^{-\lambda|x_3-y_3|}.
\label{scheme:hatK}
\ee
Thus we have
\be
K(x,t;y,s)=
\theta(t-s)\frac{e^{-b(t-s)}}{[4\pi\gamma(t-s)]^{3/2}}
e^{-\frac{(x-y)^2}{4\gamma(t-s)}},
\label{scheme:obtainedK}
\ee
where $\theta(t)$ is the Heaviside step function, i.e., $\theta=1$ for $t\ge0$ and $\theta=0$ for $t<0$. Finally, from the argument in Appendix \ref{secC}, we see $\hat{G}=2\hat{K}$ and obtain
\be
\begin{array}{ll}
G(x,t;y_1,y_2,s)
&=
2K(x,t;y,s)
\\
&=
\theta(t-s)\frac{2e^{-b(t-s)}}{[4\pi\gamma(t-s)]^{3/2}}
e^{-\frac{(x_1-y_1)^2+(x_2-y_2)^2+x_3^2}{4\gamma(t-s)}},
\end{array}
\label{scheme:Gfunc}
\ee
where we put $y_3=0$.

\subsubsection{Slab domain}

In the case of the slab domain of width $L$, we set $\Omega=\{x\in\Rm^3;\,0<x_3<L\}$. The Poisson kernel $G$ satisfies
\[
\left\{\begin{array}{lll}
\left(\pp_t-\gamma\Delta+b\right)G=0,
&\quad(x,t)\in\Omega_T,
\\
\gamma\pp_{\nu}G=\delta(x_1-y_1)\delta(x_2-y_2)\delta(t-s),
&\quad x_3=0,\;(x_1,x_2)\in\Rm^2,\;t\in(0,T),
\\
\pp_{\nu}G=0,
&\quad x_3=L,\;(x_1,x_2)\in\Rm^2,\;t\in(0,T),
\\
G=0,
&\quad x\in\Omega,\quad t=0.
\end{array}\right.
\]
Using an argument similar to Appendix \ref{secC}, we can move the boundary source to the source term in the diffusion equation as $(\pp_t-\gamma\Delta+b)G=f(x_3)\delta(x_1-y_1)\delta(x_2-y_2)\delta(t-s)$ with the boundary condition $\pp_{\nu}G=0$ at $x_3=0,L$, where $f(x_3)=\delta(x_3)$. Then we can extend $f(x_3)$ as an even $2L$-periodic function by setting $F(x_3)=f(x_3-2mL)$ for $2mL<x_3\le(2m+1)L$ and $F(x_3)=f(2(m+1)L-x_3)$ for $(2m+1)L\le x_3<2(m+1)L$, where $m=0,\pm1,\pm2,\cdots$. We have
\[
\begin{array}{ll}
G(x,t;y_1,y_2,s)
&=
\int_{-\infty}^{\infty}K(x,t;y_1,y_2,\xi,s)F(\xi)\,d\xi
\\
&=
\sum_{m=-\infty}^{\infty}\left(\int_{2mL}^{(2m+1)L}+\int_{(2m+1)L}^{2(m+1)L}
\right)K(x,t;y_1,y_2,\xi,s)F(\xi)\,d\xi
\\
&=
\sum_{m=-\infty}^{\infty}
\left[K(x,t;y_1,y_2,2mL,s)+K(x,t;y_1,y_2,2(m+1)L,s)\right],
\end{array}
\]
where $K(x,t;y_1,y_2,\xi)$ is given by replacing $x_3$ by $x_3-\xi$ in the previous $K(x,t;y,s)$ with $y_3=0$. Thus in this case, we obtain
\[
\begin{array}{ll}
G(x,t;y_1,y_2,s)
&=
2\sum_{m=-\infty}^{\infty}K(x,t;y_1,y_2,2mL,s)
\\
&=
\theta(t-s)\frac{2e^{-b(t-s)}}{[4\pi\gamma(t-s)]^{3/2}}
e^{-\frac{(x_1-y_1)^2+(x_2-y_2)^2}{4\gamma(t-s)}}
\sum_{m=-\infty}^{\infty}e^{-\frac{(x_3-2mL)^2}{4\gamma(t-s)}}.
\end{array}
\]

\subsection{Born sequence and Poisson kernel}

Let us consider $v_j$ ($j=0,1,\cdots$) introduced in (\ref{bornseries}). The $(n+1)$th term $v_n$ of the Born series satisfies
\[
\left\{\begin{array}{lll}
\left(\pp_t-\gamma\Delta+b\right)v_n=0,
&\qquad(x,t)\in\Omega_T,
\\
\gamma\pp_{\nu}v_n=-\beta v_{n-1},
&\qquad(x,t)\in\pp\Omega_T,
\\
v_n=0,
&\qquad x\in\Omega,\quad t=0,
\end{array}\right.
\]
for $n=1,2,\cdots$. The initial term is given by $v_0=u_0$. Using the Poisson kernel (\ref{scheme:Gfunc}) we have
\be
v_n(x,t)=-\beta\int_{\pp\Omega_T}G(x,t;y_1,y_2,s)
v_{n-1}(y_1,y_2,0,s)\,dy_1dy_2ds.
\label{borniteration}
\ee

We then compute $u$ as the limit of the Born sequence.
\[
u=\lim_{n\to\infty}u_n,\qquad u_n=v_0+v_1+\cdots+v_n.
\]

\section{Half space case}
\label{half}

In this section, we consider the diffusion equation (\ref{scheme:DE0}) in the half space. That is, we take $\Omega=\Rm^3_+$ and $\pp\Omega=\Rm^2$. Let
\[
g(x_1,x_2,t)=\delta(x_1)\delta(x_2)\delta(t-t_0),\,\,t_0>0.
\]
Then, by using the Poisson kernel (\ref{scheme:Gfunc}), we have
\[
\begin{array}{ll}
v_0(x,t)
&=
G(x,t;0,0,t_0),
\\
v_n(x,t)
&=
-2\beta\int_0^t\int_{\Rm^2}\frac{e^{-b(t-s)}}{[4\pi\gamma(t-s)]^{3/2}}
e^{-\frac{(x_1-y_1)^2+(x_2-y_2)^2+x_3^2}{4\gamma(t-s)}}
v_{n-1}(y_1,y_2,0,s)\,dy_1dy_2ds.
\end{array}
\]
Let us introduce $w_n(x_3,t)$ ($n=0,1,2,\cdots$) as
\[
v_n(x,t)=
\frac{e^{-b(t-t_0)}}{t-t_0}e^{-\frac{x_1^2+x_2^2}{4\gamma(t-t_0)}}w_n(x_3,t).
\]
From the definition of $G$, we have
\[
w_0(x_3,t)=
\theta(t-t_0)\frac{1}{4(\pi\gamma)^{3/2}\sqrt{t-t_0}}
e^{-\frac{x_3^2}{4\gamma(t-t_0)}}.
\]
We have the following recurrence relation for $w_n$ ($n\ge1$).
\be
w_n(x_3,t)=
\frac{-\beta}{\sqrt{\pi\gamma}}\int_0^t
\frac{w_{n-1}(0,s)}{\sqrt{t-s}}e^{-\frac{x_3^2}{4\gamma(t-s)}}\,ds.
\label{recurvn}
\ee

\begin{lem}
From the recurrence relation (\ref{recurvn}) we can show that
\[
\|w_n\|_{L^1((0,T);L^{\infty}((0,\infty)))}\le
\frac{\beta}{2}\sqrt{\frac{\pi}{\gamma}}
\|w_{n-1}\|_{L^1((0,T);L^{\infty}((0,\infty)))},
\]
where $L^{\infty}((0,\infty))$ is the set of all bounded measurable function defined in $(0,\infty)$ and $L^1((0,T);L^{\infty}((0,\infty)))$ is the set of all $L^{\infty}((0,\infty))$ valued functions which are integrable over $(0,T)$ with respect to the norm of $L^{\infty}((0,\infty))$.
\end{lem}

\begin{proof}
We note that
\[
w_n(0,t)=\frac{-\beta}{\sqrt{\pi\gamma}}\int_0^t
\frac{w_{n-1}(0,s)}{\sqrt{t-s}}\,ds.
\]
This relation implies that if $w_{n-1}(0,s)$ does not change the sign, i.e., $w_{n-1}(0,s)\ge0$ for all $s\in(0,T)$ or $w_{n-1}(0,s)\le0$ for all $s\in(0,T)$, then $w_n(0,t)$ does not change the sign neither on $(0,T)$. Indeed, we see by induction that the sign of $w_n(0,t)$ remains the same on $(0,T)$ for all $n=0,1,\cdots$ since $w_0(0,t)$ is nonnegative on $(0,T)$.

Keeping the above fact in mind, we have
\[
|w_n(0,t)|=
\frac{\beta}{\sqrt{\pi\gamma}}\left|\int_0^t\frac{w_{n-1}(0,s)}{\sqrt{t-s}}
\,ds\right|=
\frac{\beta}{\sqrt{\pi\gamma}}\int_0^t\frac{|w_{n-1}(0,s)|}{\sqrt{t-s}}\,ds.
\]
Hence,
\[
\begin{array}{ll}
\int_0^T|w_n(0,t)|2\sqrt{T-t}\,dt
&=
\frac{\beta}{\sqrt{\pi\gamma}}\int_0^T2\sqrt{T-t}\int_0^t
\frac{|w_{n-1}(0,s)|}{\sqrt{t-s}}\,dsdt
\\
&=
\frac{\beta}{\sqrt{\pi\gamma}}\int_0^T|w_{n-1}(0,s)|\int_s^T
\frac{2\sqrt{T-t}}{\sqrt{t-s}}\,dtds
\\
&=
\frac{\beta}{\sqrt{\pi\gamma}}\int_0^T|w_{n-1}(0,s)|\pi(T-s)\,ds
\\
&\le
\frac{\beta}{\sqrt{\pi\gamma}}\int_0^T|w_{n-1}(0,s)|\pi\sqrt{T-s}\,ds,
\end{array}
\]
where we used $\sqrt{T-s}\ge t-s$. Noting that $2\sqrt{T-s}=\int_s^T(1/\sqrt{t-s})\,dt$, we obtain
\[
\int_0^T|w_n(0,s)|\int_s^T\frac{1}{\sqrt{t-s}}\,dtds\le
\frac{\beta}{2}\sqrt{\frac{\pi}{\gamma}}\int_0^T|w_{n-1}(0,s)|
\int_s^T\frac{1}{\sqrt{t-s}}\,dtds.
\]
The above integrals can be rewritten as
\[
\int_0^T\int_0^t\frac{|w_n(0,s)|}{\sqrt{t-s}}\,dsdt\le
\frac{\beta}{2}\sqrt{\frac{\pi}{\gamma}}\int_0^T\int_0^t
\frac{|w_{n-1}(0,s)|}{\sqrt{t-s}}\,dsdt.
\]
Therefore,
\[
\int_0^T\left|\int_0^t\frac{w_n(0,s)}{\sqrt{t-s}}\,ds\right|\,dt\le
\frac{\beta}{2}\sqrt{\frac{\pi}{\gamma}}\int_0^T\left|\int_0^t
\frac{w_{n-1}(0,s)}{\sqrt{t-s}}\,ds\right|\,dt.
\]
This means we have
\[
\int_0^T\|\int_0^t\frac{w_n(0,s)}{\sqrt{t-s}}
e^{-\frac{x_3^2}{4\gamma(t-s)}}\,ds\|_{L^{\infty}((0,\infty))}\,dt
\le
\frac{\beta}{2}\sqrt{\frac{\pi}{\gamma}}\int_0^T\|\int_0^t
\frac{w_{n-1}(0,s)}{\sqrt{t-s}}e^{-\frac{x_3^2}{4\gamma(t-s)}}\,ds
\|_{L^{\infty}((0,\infty))}\,dt.
\]
Thus the proof is complete.
\end{proof}

Thus the series $\sum_{n=0}^{\infty}w_n$ and $\sum_{n=0}^{\infty}u_n$ converge if
\be
\beta<2\sqrt{\frac{\gamma}{\pi}}.
\label{half:betacondition}
\ee
As we will see below from the explicit calculation of $w_n$, indeed, the series converges for any $\beta$ (see Remark \ref{half:doublefactorial}).

Explicit expressions of $w_n(x_3,t)$ are available as follows. For $n\ge1$, the functions $w_n(x_3,t)$ satisfy
\begin{equation}
\begin{array}{ll}
w_n(x_3,t)\\
\quad=\frac{(-\beta)^n}{4(\pi\gamma)^{(n+3)/2}}\int_{t_0}^t\int_{t_0}^{t_n}\cdots
\int_{t_0}^{t_2}\frac{e^{-x_3^2/[4\gamma(t-t_n)]}}
{\sqrt{(t-t_n)(t_n-t_{n-1})\cdots(t_2-t_1)(t_1-t_0)}}\,dt_1\cdots dt_n
\\
\quad=
\frac{(-\beta)^n(t-t_0)^{(n-1)/2}}{4(\pi\gamma)^{(n+3)/2}}
\left[\prod_{j=1}^{n-1}\int_0^1\frac{s^{\frac{j}{2}-1}}{\sqrt{1-s}}\,ds\right]
\int_0^1\frac{s^{\frac{n}{2}-1}}{\sqrt{1-s}}
e^{-x_3^2/[4\gamma(t-t_0)(1-s)]}\,ds
\\
\quad=
\frac{(-\beta)^n(t-t_0)^{(n-1)/2}}{4(\pi\gamma)^{(n+3)/2}}
\left[\prod_{j=1}^{n-1}B\left(\frac{j}{2},\frac{1}{2}\right)\right]
e^{-\zeta^2}\int_0^{\infty}e^{-\zeta^2s}s^{\frac{n}{2}-1}
(1+s)^{-\frac{n+1}{2}}\,ds
\\
\quad=
\frac{(-\beta)^n(t-t_0)^{(n-1)/2}}{4(\pi\gamma)^{(n+3)/2}}
\frac{2^{\left\lfloor\frac{n-1}{2}\right\rfloor}
\pi^{\left\lfloor\frac{n}{2}\right\rfloor}}{(n-2)!!}
\\
\quad\times
\left[B\left(\frac{n}{2},\frac{1}{2}\right)\,
{_1}F_1\left(\frac{1-n}{2},\frac{1}{2};-\zeta^2\right)-2\sqrt{\pi}\zeta\,
{_1}F_1\left(1-\frac{n}{2},\frac{3}{2};-\zeta^2\right)\right],
\label{half:vnexplicit}
\end{array}
\end{equation}
where the floor function $\lfloor\cdot\rfloor$ is defined such that $\lfloor x\rfloor$ ($x\in\Rm$) denotes the largest integer which does not exceed $x$, and double factorials $n!!=n\cdot(n-2)\cdot(n-4)\cdots$ are defined with $(-1)!!=0!!=1$. Here, $\zeta=x_3/\sqrt{4\gamma(t-t_0)}$, $B$ is the beta function, and ${_1}F_1$ is the Kummer confluent hypergeometric function of the first kind. See Appendix \ref{specialfunctions} for the computation of $w_n(x_3,t)$. In particular, we have
\[
w_n(0,t)=\frac{(-\beta)^n(t-t_0)^{(n-1)/2}}{4(\pi\gamma)^{(n+3)/2}}
\frac{2^{\left\lfloor\frac{n}{2}\right\rfloor}
\pi^{\left\lfloor\frac{n+1}{2}\right\rfloor}}{(n-1)!!}.
\]

Finally, we arrive at
\be
\begin{array}{ll}
u(x,t)
&=
v_0(x,t)+v_1(x,t)+\cdots
\\
&=
\frac{e^{-b(t-t_0)}}{t-t_0}e^{-\frac{x_1^2+x_2^2}{4\gamma(t-t_0)}}
\left[w_0(x_3,t)+w_1(x_3,t)+\cdots\right].
\end{array}
\label{half:final}
\ee

\begin{rmk}
\label{half:doublefactorial}
Due to the double factorial $(n-2)!!$ in the denominator of each $n$th term of $w_0+w_1+\cdots$ in (\ref{half:vnexplicit}), clearly $\left|w_n/w_{n-1}\right|<1$ for sufficiently large $n$. Therefore the series $\sum_{n=0}^{\infty}w_n$ and thus $\sum_{n=0}^{\infty}v_n$ locally uniformly converge regardless of the value of $\beta$.
\end{rmk}

\section{Numerical calculation}
\label{numerical}

For numerical calculation, we set $t_0$ to be zero (cf., (\ref{ana:set0})), and also set $x_1=x_2=0$. Then the $n$th Born approximation for (\ref{half:final}) is written as
\be
u_n(x_3,t)=\frac{e^{-bt}}{t}\sum_{j=0}^nw_j(x_3,t).
\label{numerical:solBorn}
\ee

\begin{figure}[ht]
\includegraphics[width=0.45\textwidth]{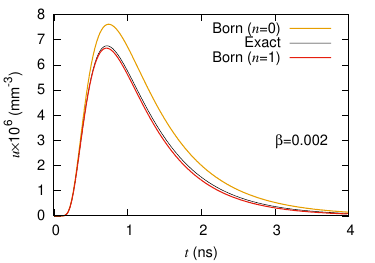}
\includegraphics[width=0.45\textwidth]{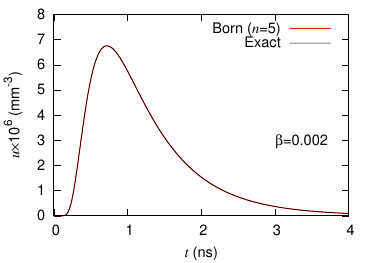}
\caption{
The energy density $u$ is plotted at $x_3=20\,{\rm mm}$ as a function of $t$ for $\beta=0.002\,{\rm mm}/{\rm ps}$. (Left) From the top, $u_0(x_3,t)$, $u(x_3,t)$, and $u_1(x_3,t)$ are shown. (Right) We plot $u_5(x_3,t)$ and $u(x_3,t)$. The two curves are almost identical.
}
\label{fig:beta0.002}
\end{figure}

\begin{figure}[ht]
\includegraphics[width=0.45\textwidth]{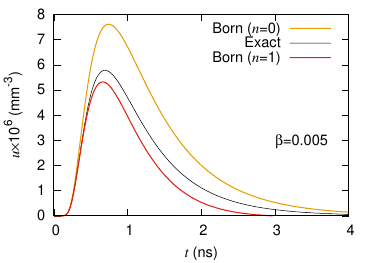}
\includegraphics[width=0.45\textwidth]{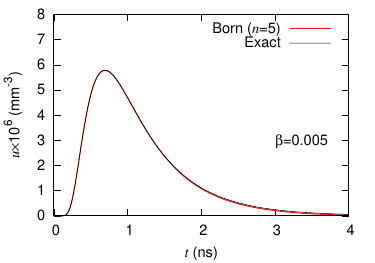}
\caption{
The energy density $u$ is plotted at $x_3=20\,{\rm mm}$ as a function of $t$ for $\beta=0.005\,{\rm mm}/{\rm ps}$. (Left) From the top, $u_0(x_3,t)$, $u(x_3,t)$, and $u_1(x_3,t)$ are shown. (Right) We plot $u_5(x_3,t)$ and $u(x_3,t)$. Two energy densities for $u(x_3,t)$ and $u_5(x_3,t)$ are almost indistinguishable.
}
\label{fig:beta0.005}
\end{figure}

\begin{figure}[ht]
\includegraphics[width=0.45\textwidth]{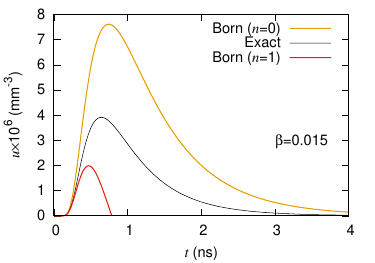}
\includegraphics[width=0.45\textwidth]{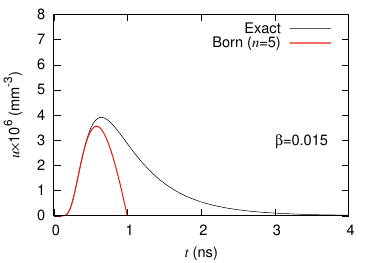}
\caption{
The energy density $u$ is plotted at $x_3=20\,{\rm mm}$ as a function of $t$ for $\beta=0.015\,{\rm mm}/{\rm ps}$. (Left) From the top, $u_0(x_3,t)$, $u(x_3,t)$, and $u_1(x_3,t)$ are shown. (Right) From the top, $u(x_3,t)$ and $u_5(x_3,t)$ are shown.
}
\label{fig:beta0.015a}
\end{figure}

\begin{figure}[ht]
\includegraphics[width=0.45\textwidth]{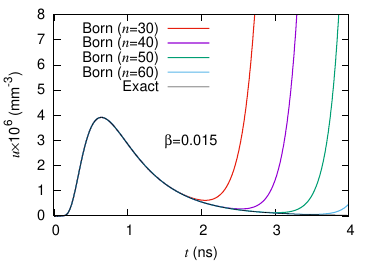}
\includegraphics[width=0.45\textwidth]{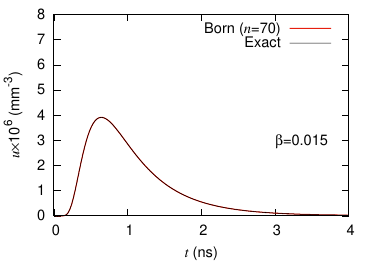}
\caption{
Same as Fig.~\ref{fig:beta0.015a} but the $30$th through $70$th Born approximations are presented. (Left) From the top to the bottom, $u_{30}(x_3,t)$, $u_{40}(x_3,t)$, $u_{50}(x_3,t)$, $u_{60}(x_3,t)$, and $u(x_3,t)$ are shown. The curves show an excellent agreement except their tails. (Right) The results for $u_{70}(x_3,t)$ and $u(x_3,t)$ are shown. The case of $n=70$ gives a numerically exact result.
}
\label{fig:beta0.015b}
\end{figure}

Let us compare $u_n(x_3,t)$ in (\ref{numerical:solBorn}) and $u(x_3,t)$ in (\ref{ana:solexact}) using the parameter values given in (\ref{ana:para}). As is seen in Fig.~\ref{fig:beta0.002}, $n=1$ is already a good approximation when $\beta=0.002$. In Fig.~\ref{fig:beta0.005}, we set $\beta=0.005$. We see that the energy density from the Born approximation of $n=5$ becomes indistinguishable from the exact solution. In Figs.~\ref{fig:beta0.015a} and \ref{fig:beta0.015b}, we set $\beta=0.015$. Since the value of $\beta$ is larger, we need to take more terms. We arrive at the numerically exact result for $n=70$.

The left panel of Fig.~\ref{fig:beta0.015b} suggests how the necessary number of terms $n$ can be determined. Since results from different $n$ agree for short time, we should use $n$ such that curves for terms greater than or equal to $n$ agree until $t=T$. Although it is not easy to know the optimal $n$ a priori, we can find such $n$ by trying several $n$'s.

Numerical calculation was done by Mathematica using a single Intel Core i5 (2.9 GHz). The computation time for $\beta=0.002$, $n=5$ in Fig.~\ref{fig:beta0.002} and $\beta=0.005$, $n=5$ in Fig.~\ref{fig:beta0.005} were 0.4 sec whereas for $\beta=0.015$ in Fig.~\ref{fig:beta0.015b}, the cases $n=30,40,50,60,70$ required 2.4, 3.3, 4.8, 6.0, and 7.7 sec, respectively. The present formulation is beneficial when the Robin boundary condition with small $\beta$ is considered. If we suppose that the diffusion approximation holds on the boundary and assume the diffuse surface reflection, we have $\beta=c/(2A)$, where $c$ is the speed of light in the medium and $A=(1+r_d)/(1-r_d)$ with the internal reflection $r_d$ \cite{Groenhuis-Ferwerda-TenBosch83}. Let us suppose the reflective index outside the medium is unity. The refractive indices $n=1.7$, $2.3$, $2.9$ correspond to $\beta=0.016$, $0.0053$, and $0.0020$, respectively.

\section{Concluding remarks}
\label{concl}

In Sec.~\ref{numerical}, we considered the half space case and validated our approach of applying the Born series for boundary conditions. The comparison of Figs.~\ref{fig:ebc} and \ref{fig:beta0.002} suggests that the present approach provides an efficient alternative formula for small $\beta$ when the approximation with the extrapolated boundary condition does not work well. It is important for our formulation that the solution for the Neumann boundary condition has a simple explicit form. We explored the Poisson kernel in the half space and slab domain in Sec.~\ref{scheme}. Applying the present strategy to other geometries is an interesting future problem.

\section*{Acknowledgment}
The first author acknowledges support from Grant-in-Aid for Scientific Research (17H02081, 17K05572, and 18K03438) of the Japan Society for the Promotion of Science (JSPS) and from the JSPS A3 foresight program: Modeling and Computation of Applied Inverse Problems. This work was also supported by HUSM Grant-in-Aid granted to the first author. The second author was partially supported by Grant-in-Aid for Scientific Research (15K21766 and 15H05740) of the Japan Society for the Promotion of Science doing the research of this paper.

\appendix

\section{Exact solutions}
\label{alt}

Here, we compute the Poisson kernel for the three space dimensional half space, i.e., we solve (\ref{ana:de}). See also \cite{Hielscher-Jacques-Wang-Tittel95,Yosida-Ito}.

Define
\[
\phi:=u-G
\]
with $G$ is given in (\ref{scheme:Geq}). Then $\phi$ satisfies
\[
\left\{\begin{array}{lll}
\left(\pp_t-\gamma\Delta+b\right)\phi=0,&\qquad x\in\Omega,\quad t>0,
\\
\gamma\pp_{\nu}\phi+\beta\phi=-\beta G,
&\qquad x\in\pp\Omega,\quad t>0,
\\
\phi=0,&\qquad x\in\Omega,\quad t=0.
\end{array}\right.
\]
Recall that $\lambda$ is given in (\ref{scheme:deflambda}). The Laplace-Fourier transform given by
\[
\hat{\phi}(x_3;p,q;y_1,y_2,s)=\int_0^{\infty}\int_{\Rm^2}e^{-pt}
e^{-i(q_1x_1+q_2x_2)}\phi(x,t;y_1,y_2,s)\,dx_1dx_2dt,
\]
satisfies
\[
\left\{\begin{array}{lll}
-\frac{d^2}{dx_3^2}\hat{\phi}+\lambda^2\hat{\phi}=0,
&\qquad x_3>0,
\\
-\gamma\frac{d\hat{\phi}}{dx_3}+\beta\hat{\phi}=-\beta \hat{G},
&\qquad x_3=0.
\end{array}\right.
\]

On the other hand we have \eqref{scheme:hatK} and $\hat{G}$ can be given by
\begin{equation}\label{hat G}
\hat{G}(x_3;p,q;y_1,y_2,s)=\frac{1}{\lambda\gamma}
e^{-ps-i(q_1y_1+q_2y_2)}e^{-\lambda x_3}.
\end{equation}
Here we used the relation $\hat{G}=2\hat{K}$ (see Appendix \ref{secC}). Hence,
\be
\hat{\phi}(x_3)=
\frac{-\beta}{\beta+\lambda\gamma}\hat{G}(0)e^{-\lambda x_3}=
\frac{-\beta}{\lambda\gamma(\beta+\lambda\gamma)}
e^{-ps}e^{-i(q_1y_1+q_2y_2)}e^{-\lambda x_3}.
\label{alt:hatphi}
\ee
Further from the relation
\[
\frac{d}{dx_3}\hat{\phi}(x_3)=
\frac{2\beta}{\gamma}\hat{K}(x_3,0)+\frac{\beta}{\gamma}\hat{\phi}(x_3)
\]
which can be readily verified using (\ref{scheme:hatK}) and (\ref{alt:hatphi}), we have
\[
\hat{\phi}(x_3)=-\frac{2\beta}{\gamma}\int_{x_3}^{\infty}
e^{\frac{\beta}{\gamma}(x_3-\xi)}\hat{K}(\xi,0)\,d\xi.
\]
Thus we arrive at the following solution.
\[
\begin{array}{ll}
u(x,t)
&=
G(x,t;y_1,y_2,t_0)
-\frac{2\beta}{\gamma}\int_{x_3}^{\infty}
e^{\frac{\beta}{\gamma}(x_3-\xi)}K(x_1,x_2,\xi,t;y_1,y_2,0,t_0)\,d\xi
\\
&=
\theta(t-t_0)\frac{2e^{-b(t-t_0)}}{[4\pi\gamma(t-t_0)]^{3/2}}
e^{-\frac{(x_1-y_1)^2+(x_2-y_2)^2+x_3^2}{4\gamma(t-t_0)}}
\\
&-
\theta(t-t_0)\frac{\beta e^{-b(t-t_0)}}{4\pi\gamma^2(t-t_0)}
e^{-\frac{(x_1-y_1)^2+(x_2-y_2)^2}{4\gamma(t-t_0)}}
e^{\frac{\beta}{\gamma}\left(x_3+\beta(t-t_0)\right)}\mathop{\mathrm{erfc}}
\left(\frac{x_3+2\beta(t-t_0)}{\sqrt{4\gamma(t-t_0)}}\right),
\end{array}
\]
where $G$ is given in (\ref{scheme:Gfunc}).

\section{An interpretation of transient boundary point source}\label{secC}

By using the advantage of the simple geometry for $\Omega$,  we will explain more explicity than given before in Sec.~\ref{general} how the solution $u$ of \eqref{ana:de} with a transient boundary point source can be obtain as a limit of the solution $u_{\epsilon}$ of the following initial boundary value problem with a transient point source:
\[
\left\{\begin{array}{lll}
\left(\pp_t-\gamma\Delta+b\right)u_{\epsilon}=g\delta(x_3-\epsilon),
&\qquad(x,t)\in\Omega_T,
\\
\gamma\pp_{\nu}u_{\epsilon}+\beta u_{\epsilon}=0,
&\qquad(x,t)\in\pp\Omega_T,
\\
u_{\epsilon}=0,
&\qquad x\in\Omega,\quad t=0
\end{array}\right.
\]
with $g=\delta(x_1-y_1)\delta(x_2-y_2)\delta(t-s)$. We prepare the following $G_{\epsilon}$.
\[
\left\{\begin{array}{lll}
\left(\pp_t-\gamma\Delta+b\right)G_{\epsilon}=g\delta(x_3-\epsilon),
&\qquad(x,t)\in\Omega_T,
\\
\gamma\pp_{\nu}G_{\epsilon}=0,
&\qquad(x,t)\in\pp\Omega_T,
\\
G_{\epsilon}=0,
&\qquad x\in\Omega,\quad t=0
\end{array}\right.
\]
with the same $g$ as above. Similar to the calculation in Appendix \ref{alt}, let us consider $u_{\epsilon}$ in the form (\ref{secC:defphiepsilon}) below.
\be
\phi_{\epsilon}=u_{\epsilon}-G_{\epsilon}.
\label{secC:defphiepsilon}
\ee
Here $\phi_{\epsilon}$ satisfies
\[
\left\{\begin{array}{lll}
\left(\pp_t-\gamma\Delta+b\right)\phi_{\epsilon}=0,
&\qquad(x,t)\in\Omega_T,
\\
\gamma\pp_{\nu}\phi_{\epsilon}+\beta \phi_{\epsilon}=
-\beta G_{\epsilon},
&\qquad(x,t)\in\pp\Omega_T,
\\
\phi_{\epsilon}=0,
&\qquad x\in\Omega,\quad t=0.
\end{array}\right.
\]
We obtain
\[
\hat{\phi}_{\epsilon}(x_3)=
\frac{-\beta}{\beta+\lambda\gamma}\hat{G}_{\epsilon}(0)e^{-\lambda x_3}.
\]
Here,
\[
\hat{G}_{\epsilon}(x_3)=
\int_0^{\infty}\int_{\Rm^2}e^{-pt}e^{-i(q_1x_1+q_2x_2)}
G_{\epsilon}(x,t;y_1,y_2,\epsilon,s)\,dx_1dx_2dt.
\]
We note that
\[
\begin{array}{ll}
G_{\epsilon}(x,t;y_1,y_2,\epsilon,s)=
K(x,t;y_1,y_2,\epsilon,s)+K(x,t;y_1,y_2,-\epsilon,s),
\end{array}
\]
where $K$ is given in (\ref{scheme:obtainedK}). Therefore we obtain
\[
\hat{G}_{\epsilon}(x_3)=
\frac{1}{2\lambda\gamma}e^{-ps}e^{-i(q_1y_1+q_2y_2)}\left(
e^{-\lambda|x_3-\epsilon|}+e^{-\lambda|x_3+\epsilon|}\right),
\]
where we used (\ref{scheme:hatK}). In the limit we have $\lim_{\epsilon\to0}\hat{G}_{\epsilon}=\hat{G}$, which is given in (\ref{hat G}). Thus we arrive at
\[
\hat{\phi}_{\epsilon}(x_3)=
\frac{-\beta}{\lambda\gamma(\beta+\lambda\gamma)}e^{-ps}
e^{-i(q_1y_1+q_2y_2)}e^{-\lambda(x_3+\epsilon)}.
\]
We see that $\lim_{\epsilon\to0}\hat{\phi}_{\epsilon}=\hat{\phi}$, which is given in (\ref{alt:hatphi}). Thus we can directly see that the distribution $u_{\epsilon}$ converges to the distribution $u$ as $\epsilon\to0$.

\section{Special functions}
\label{specialfunctions}

By using the formulae
\[
B\left(a,\frac{1}{2}\right)B\left(a+\frac{1}{2},\frac{1}{2}\right)=
\frac{\pi}{a},\qquad
B\left(\frac{1}{2},\frac{1}{2}\right)=\pi,
\]
we have
\[
B\left(\frac{1}{2},\frac{1}{2}\right)B\left(1,\frac{1}{2}\right)
B\left(\frac{3}{2},\frac{1}{2}\right)\cdots
B\left(\frac{n-1}{2},\frac{1}{2}\right)
=\left\{\begin{aligned}
\frac{2^{\frac{n}{2}-1}\pi^{\frac{n}{2}}}{(n-2)!!}&\quad\mbox{($n$ even)},
\\
\frac{2^{\frac{n-1}{2}}\pi^{\frac{n-1}{2}}}{(n-2)!!}&\quad\mbox{($n$ odd)}.
\end{aligned}\right.
\]
Moreover,
\[
B\left(\frac{n}{2},\frac{1}{2}\right)=
\frac{\Gamma\left(\frac{n}{2}\right)\Gamma\left(\frac{1}{2}\right)}
{\Gamma\left(\frac{n+1}{2}\right)},
\]
where $\Gamma(\frac{1}{2})=\sqrt{\pi}$.

Now recall the Kummer confluent hypergeometric function of the first kind is given by
\[
{_1}F_1(a,b;z)=M(a,b;z)=\sum_{n=0}^{\infty}\frac{(a)_n}{(b)_nn!}z^n
=1+\frac{a}{b}z+\frac{a(a+1)}{b(b+1)2!}z^2+\cdots.
\]
Then we have
\[
\begin{array}{ll}
{_1}F_1\left(0,\frac{1}{2};-z\right)&=1,
\\
{_1}F_1\left(-\frac{1}{2},\frac{1}{2};-z\right)&=
e^{-z}+\sqrt{\pi z}\mathop{\mathrm{erf}}(\sqrt{z}),
\\
{_1}F_1\left(-1,\frac{1}{2};-z\right)&=1+2z,
\\
{_1}F_1\left(-\frac{3}{2},\frac{1}{2};-z\right)&=
(1+z)e^{-z}+\sqrt{\pi z}\left(z+\frac{3}{2}\right)
\mathop{\mathrm{erf}}(\sqrt{z}),
\end{array}
\]
and
\[
\begin{array}{ll}
{_1}F_1\left(\frac{1}{2},\frac{3}{2};-z\right)&=
\frac{1}{2}\sqrt{\frac{\pi}{z}}\mathop{\mathrm{erf}}(\sqrt{z}),
\\
{_1}F_1\left(0,\frac{3}{2};-z\right)&=1,
\\
{_1}F_1\left(-\frac{1}{2},\frac{3}{2};-z\right)&=
\frac{e^{-z}}{2}+\frac{\sqrt{\pi z}}{2}\left(1+\frac{1}{2z}\right)
\mathop{\mathrm{erf}}(\sqrt{z}),
\\
{_1}F_1\left(-1,\frac{3}{2};-z\right)&=1+\frac{2}{3}z,
\end{array}
\]
where
\[
\mathop{\mathrm{erf}}(\sqrt{z})=
\frac{2}{\sqrt{\pi}}\int_0^{\sqrt{z}}e^{-t^2}\,dt.
\]

We close this Appendix \ref{secC} by giving some miscellaneous facts on hypergoemetic function and error function which are useful for computing the Poisson kernel numerically. Besides the hypergeometric function given above explicitly, other hypergeometric functions can be recursively computed using the following recurrence relation:
\[
{_1}F_1(a-1,b;z)=
\frac{2a-b+z}{a-b}{_1}F_1(a,b;z)-\frac{a}{a-b}{_1}F_1(a+1,b;z).
\]
The following form is convenient to numerically evaluate the error function:
\[
\mathop{\mathrm{erf}}(\xi)=
\frac{2}{\sqrt{\pi}}\sum_{n=0}^{\infty}
\frac{(-1)^n\xi^{2n+1}}{n!(2n+1)}=
\frac{2}{\sqrt{\pi}}e^{-\xi^2}\sum_{n=0}^{\infty}
\frac{2^n\xi^{2n+1}}{(2n+1)!!}.
\]


\begin{thebibliography}{99}

\bibitem{Arridge99}
Arridge, S. R.,
``Optical tomography in medical imaging,''
{\it Inverse Problems} {\bf 15}, R41--R93 (1999).

\bibitem{Arridge-Cope-Delpy92}
Arridge, S. R., Cope, M., and Delpy, D. T.,
``The theoretical basis for the determination of optical pathlengths in tissue: temporal and frequency analysis,''
{\it Phys. Med. Biol.} {\bf 37}, 1531--1560 (1992).

\bibitem{Arridge-Schotland09}
Arridge, S. R. and Schotland, J. C.,
``Optical tomography: forward and inverse problems,''
{\it Inverse Problems} {\bf 25}, 123010 (2009).

\bibitem{Ayyalasomayajula-Yalavarthy13}
Ayyalasomayajula, K. R. and Yalavarthy, P. K.,
``Analytical solutions for diffuse fluorescence spectroscopy/imaging in biological tissues. Part I: zero and extrapolated boundary conditions,''
{\it J. Opt. Soc. Am. A} {\bf 30}, 537--552 (2013).

\bibitem{Boas-etal2001}
Boas, D. A., Brooks, D. H., Miller, E. L., DiMarzio, C. A., Kilmer, M., Gaudette, R. J., and Zhang, Q.,
``Imaging the body with diffuse optical tomography,''
{\it IEEE Signal Processing Magazine} {\bf 18}, 57--75 (2001).

\bibitem{Duderstadt-Hamilton}
Duderstadt, J. J. and Hamilton, L. J.,
{\it Transport Theory}
(New York: John Wiley \& Sons, 1976).

\bibitem{Groenhuis-Ferwerda-TenBosch83}
Groenhuis, R. A. J., Ferwerda, H. A., and Ten Bosch, J. J.,
``Scattering and absorption of turbid materials determined from reflection measurements. 1: Theory,''
{\it Appl. Opt. } {\bf 22}, 2456--2462 (1983).

\bibitem{Hielscher-Jacques-Wang-Tittel95}
Hielscher, A. H., Jacques, S. L., Wang, L., and Tittel, F. K.,
``The influence of boundary conditions on the accuracy of diffusion theory in time-resolved reflectance spectroscopy of biological tissues,''
{\it Phys. Med. Biol.} {\bf 40}, 1957--1975 (1995).

\bibitem{Hormander}
H$\ddot{\rm o}$rmander, L., {\it The analysis of linear partial differential operators, vol. I} (New York, Berlin, Tokyo: Springer 1983).

\bibitem{Hoshi16}
Hoshi, Y.,
{\it Chapter 7: Hemodynamic signals in fNIRS}
Progress in Brain Research Vol.~225 153--179 
ed.~Masamoto, K., Hirase, H., and Yamada, K.
(Amsterdam: Elsevier, 2016).

\bibitem{Ishimaru78}
Ishimaru, A.,
{\it Wave Propagation and Scattering in Random Media}
(New York: Academic, 1978)

\bibitem{Kienle-etal98}
Kienle, A., Patterson, M. S., D\"{o}gnitz, N., Bays, R., Wagni\`{e}res, G., and van den Bergh, H., ``Noninvasive determination of the optical properties of two-layered turbid media,''
{\it Appl. Opt.} {\bf 37}, 779--791 (1998).

\bibitem{Konecky-etal09}
Konecky, S. D., Mazhar, A., Cuccia, D., Durkin, A. J., Schotland, J. C., and Tromberg, B. J., ``Quantitative optical tomography of sub-surface heterogeneities using spatially modulated structured light,''
{\it Opt. Exp.} {\bf 17}, 14780--14790 (2009).

\bibitem{McLean}
McLean, W.,
{\it Strongly Elliptic Systems and Boundary Integral Equations}
(Cambridge, Cambridge University Press, 2000).

\bibitem{Martelli-etal}
Martelli, F., Del Bianco, S., Ismaelli, A., and Zaccanti, G.,
{\it Light Propagation through Biological Tissue and Other Diffusive Media}
(Washington: SPIE Press, 2010).

\bibitem{NW}
Nakamura, G. and Wang, H., ``Solvability of interior transmission problem for the diffusion equation by constructing its Green function,''
to appear in {\it Journal of Inverse and Ill-posed Problems} (2019).

\bibitem{Panasyuk-etal08}
Panasyuk, G. Y., Wang, Z.-M., Schotland, J. C., and Markel, V. A., ``Fluorescent optical tomography with large data sets,''
{\it Opt. Lett.} {\bf 33}, 1744--1746 (2008).

\bibitem{Patterson-Chance-Wilson89}
Patterson, M. S., Chance, B., and Wilson, B. C.,
``Time resolved reflectance and transmittance for the noninvasive measurement of tissue optical properties,''
{\it Appl. Opt.} {\bf 28}, 2331--2336 (1989).

\bibitem{Ryzhik-Papanicolaou-Keller96}
Ryzhik, L., Papanicolaou, G., and Keller, J. B.,
``Transport equations for elastic and other waves in random media,''
{\it Wave Motion} {\bf 24}, 327--370 (1996).

\bibitem{Schweiger-etal95}
Schweiger, M., Arridge, S. R., Hiraoka, M., and Delpy, D. T.,
``The finite element method for the propagation of light in scattering media: 
Boundary and source conditions,''
{\it Med. Phys.} {\bf 22}, 1779--1792 (1995).

\bibitem{Wloka}
Wloka, J.,
{\it Partial differential equations}
(Cambridge: Cambridge University Press, 1987)

\bibitem{Yodh-Chance95}
Yodh, A. and Chance, B.,
``Spectroscopy and imaging with diffusing light,''
{\it Physics Today} {\bf 48}, 34--40 (1995).

\bibitem{Yoo-Liu-Alfano90}
Yoo, K. M., Liu, F., and Alfano, R. R.,
``When does the diffusion approximation fail to describe photon transport in 
random media?''
{\it Phys. Rev. Lett.} {\bf 64}, 2647--2650 (1990).

\bibitem{Yosida-Ito}
Yosida, K. and Ito, S.,
{\it Functional analysis and differential equations} [Japanese]
(Tokyo: Iwanami, 1976).

\end{thebibliography}
\end{document}